\documentstyle[aps,prl,multicol]{revtex}
\input{epsf.tex}
\begin{document}
\draft
\title{BCS superfluidity in ultracold gases with unequal atomic populations}
\author{R. Combescot}
\address{Laboratoire de Physique Statistique,
 Ecole Normale Sup\'erieure*,
24 rue Lhomond, 75231 Paris Cedex 05, France}
\date{Received \today}
\maketitle

\begin{abstract}
We consider the existence of a BCS superfluid phase in $^{6}$Li due 
to the pairing of two hyperfine states with unequal number of atoms. 
We show that the domain of existence for this phase will be increased to 
a very large extent in the vicinity of the threshold for collapse. This is 
due to the presence of a new phase with anisotropic order 
parameter. This phase is induced by the anisotropic part of the 
scattering, linked to the indirect interaction between atoms through 
density fluctuation exchange.
\end{abstract}
\pacs{PACS numbers :  03.75.Fi, 32.80.Pj, 67.90.+z, 74.20.Fg}
\begin{multicols}{2}
Much progress has been made recently toward the experimental 
investigation of degenerate ultracold Fermi gases. Indeed the degenerate 
regime has already been reached in  $^{40}$K by sympathetic cooling 
of two hyperfine states \cite{marcojin} whereas the possibility of 
cooling a mixture of $^{6}$Li and $^{7}$Li atoms in order to reach 
this range has been demonstrated \cite{mfsss}. In addition to a number 
of very interesting physical effects linked to phase space restriction 
\cite{marcojin,gf} one of the most fascinating prospect for 
experiment is the possibility \cite{stoofal} to reach the onset of Cooper 
pairs condensation and observe the resulting BCS superfluid, the 
equivalent of Bose Einstein condensates for Fermi gases. As we will 
see this phase is likely to present quite unusual features.

In addition to the problems raised by the use of evaporative cooling for 
Fermi gases a possible stumbling block in the path to the BCS superfluid 
is the need to achieve a near equality \cite{stoofal,bkk} between the 
number of atoms in the two hyperfine states assumed to form pairs. 
Indeed in the various schemes under investigation there is no fast 
relaxation mechanism which equalizes these two populations and the 
number of atoms in each state will be essentially conserved. Therefore 
in these ultracold atomic gases, one will necessarily have to deal with 
situations where the two populations of particles assumed to form pairs 
are not equal. Now the difference in chemical potential between these 
two populations acts as an effective field which tends to break pairs and 
the superfluid phase is destroyed when this difference is of order of the 
critical temperature. Hence this pair breaking effect becomes more important 
a problem if the critical temperature is low and, since there is presently 
some uncertainty \cite{stoofal,rc} in the value of $T_{c}$, it is clearly 
of interest to investigate this question in detail. Moreover, 
independently of this possible experimental problem, this difference in 
chemical potential is an additional control parameter in the system quite 
interesting to play with and it is worth studying its effect, all the more 
since it is not trivial.

This problem of pairing with two unequal populations has already been 
considered a long time ago in the context of superconductivity. Here 
this is an applied magnetic field which tends to make the two spin 
populations unequal. Usually the critical field is limited by orbital 
effects. However the question of the limitation of the superconducting 
phase, when these orbital effects are small, has been considered very 
early and it was pointed out by Clogston \cite{clog} and Chandrasekhar 
\cite{chand} that the standard BCS phase could, at most, resist to a 
difference in chemical potential between the two populations of the 
order of the critical temperature (the so-called paramagnetic limit). Not 
long after, Fulde and Ferrell \cite{ff} (FF), and independently Larkin 
and Ovchinnikov \cite{larkov} (LO), showed that pairing could 
somewhat adjust to the difference in chemical potential, instead of just 
resisting, by letting the pairs have a common momentum ${\bf  K}$ 
instead of having it equal to zero as for a standard superconductor. 
However the effect was actually found rather small. 
Indeed at zero temperature the standard BCS phase goes to the 
normal state by a first order transition \cite{clog} when the chemical 
potential difference $ 2 \bar{ \mu }$ is equal to $ \sqrt{2} \Delta _{0}$ 
, where $ \Delta _{0} = 1.76 T_{c} $ is the zero temperature gap for 
equal populations. The FFLO phase goes to the normal state by a 
second order transition for $  \bar{ \mu } = 0.754 \Delta _{0} $, which 
is not much beyond. Actually there is to date no undisputed observation 
of this phase in standard superconductors \cite{br}, most likely because 
it is quite sensitive to impurities. It is the purpose of the present paper to 
show that pairing can adjust even better than in the FFLO phase, and 
that this new phase can in some instances lead to quite a strong increase 
of the existence domain for the superfluid phase.

One can view the appearance of the FFLO phase in the following way. 
When the chemical potentials for the two spin populations are different, 
it becomes more costly in terms of kinetic energy to form $({\bf  k},- 
{\bf  k})$ pairs in the standard BCS way because one can not pick the 
two particles very near the Fermi surfaces, since these surfaces do not 
match due to their size difference. In the FFLO phase this problem is 
remedied by taking a nonzero total momentum which amounts to 
shifting, in momentum space, one of the Fermi surface with respect to 
the other so that they almost match on some region. But naturally 
matching gets worst on the opposite side. Nevertheless the total balance 
is barely favourable. Now one can think to improve this situation by 
making pairing stronger on the side where there is energy gain and 
weaker where there is energy loss. This means one looks for 
anisotropic pairing to find a lower energy ground state. Naturally this 
can not work if scattering is isotropic, as it will be essentially the 
case for a weakly interacting ultracold gas since p-wave scattering is 
negligible, and this will be indeed the situation in the dilute regime 
where the coupling constant $ \lambda = 2 k_{F} | a | / \pi $ is small. 
However, even if the bare scattering is isotropic, the renormalized 
interaction is not because of the existence of the Fermi surface. This 
effect is at the basis of the Kohn and Luttinger \cite{kl} paper, where 
they showed that, even with a repulsive interaction, Cooper pairs would 
necessarily form in high angular momentum. For our purpose the case 
of $ ^{6}$Li is of particular interest since the high density regime is 
bounded by an instability \cite{stoofal} occuring for $ \lambda \geq 1 $ . 
In the vicinity of this instability the effective interaction will be 
strongly anisotropic, leading to a marked increase in the existence domain 
of the superfluid phase, compared to the FFLO phase. Note that most 
experiments are likely to be done in this range since it corresponds to 
higher critical temperature. We remark also that a similar phase with
anisotropic order parameter should exist in clean superconductors with
Pauli limited upper critical field and anisotropic interaction.

Before going into the effect of anisotropic pairing, it is of interest to 
show that this is the best possible choice for pairing with unequal 
particle number. Indeed assume that we look for the most general kind 
of pairing, where $ {\bf  k} \uparrow $ is paired with $ f({\bf  k})  
\downarrow $. On one hand we want pairing to be stable against 
scattering, which means that scattering must send this pair into another 
pair $ {\bf k} ' \uparrow $ , $ f({\bf  k}') \downarrow $. On the other 
hand scattering conserves momentum which leads, for any $ {\bf  k}$ 
and ${\bf  k}'$, to $ {\bf  k}+ f({\bf  k}) =  {\bf  k}' + f({\bf  k} ') = 
{\bf  K}$ where ${\bf  K}$ is a constant. This shows that $ f({\bf  k}) 
= {\bf  K} - {\bf  k}$, that is pairs with a total momentum  ${\bf  K}$ 
is the most general solution. Note that this argument assumes 
translational invariance. This will not hold for a finite sample such as 
those obtained with trapped gases, which will induce surface effects. 
However these surface effects should be small if we do not want to 
have $T_{c}$ drastically reduced \cite{castin} (more precisely we want 
the pair size to be small compared to the sample size). So it is 
reasonable to neglect size effects in a first step. Now the common 
momentum ${\bf  K}$ produces a breaking of rotational symmetry, and 
the standard symmetry analysis leading to pairs with a given angular 
momentum (s-wave pairs in the case of  $ ^{6}$Li) is no longer valid. 
The general order parameter $ \Delta _{ \bf k}$ will depend on the 
wavevector $ {\bf  k}$. However we still have rotational invariance 
around  ${\bf  K}$ and we can classify the solutions by their angular 
momentum $m$. We will assume that the most stable pairing 
corresponds to $ m = 0 $, as it is likely to be so for a standard 
interaction. However other values of $m$, corresponding to a breaking 
of the rotational symmetry around  ${\bf  K}$, do not seem to be 
excluded from first principles and would be undoubtedly a very 
interesting situation.

We will explore now quantitatively the possibility offered by an 
anisotropic order parameter. However our purpose is more to 
demonstrate the importance of the effect than to perform an exact
calculation. This last goal would require a perfect knowledge of the 
effective interaction, which is not available. We will consider 
specifically the case of $ ^{6}$Li and we restrict ourselves to the 
determination of the T = 0 critical difference in chemical potential  $ 2 
\bar{ \mu } \equiv \mu \uparrow - \mu  \downarrow $ above which 
superfluidity disappears. Since we want to go in the high density 
regime, we need to have an expression for the effective interaction in 
this range. For this purpose we take the paramagnon model which we 
have already used for an evaluation of $T_{c}$ in the high density 
regime \cite{rc}. Actually we will not retain the full interaction of this 
model because most of the terms lead to a moderate anisotropy and they 
would produce a more complex calculation without adding much effect. 
The essential contribution of those terms is already included in the value 
$T  ^{0} _{c}$ of the critical temperature for equal populations. We 
will only retain the explicit attractive part coming from density 
fluctuations which produces near the instability a strong contribution for 
low momentum transfer and in this way lead to a strongly anisotropic 
interaction. Moreover we will for simplicity perform a weak coupling 
calculation, omitting all the frequency dependence and taking the zero 
frequency value of the interaction. To be coherent with this weak 
coupling approach we will also omit self-energy effects. This leads to 
the total effective attractive interaction $ V( {\bf  k}, {\bf  k}')$ 
\cite{rc} given by :
\begin{eqnarray}
N _{f}V( {\bf  k}, {\bf  k}') = \lambda  + \frac{1}{2} \ \frac{ \lambda 
^{2} \bar{\chi } _{0}( {\bf  q}) }{1- \lambda \bar{\chi } _{0}({\bf  
q})} 
\label{eq1}
\end{eqnarray}
with $ {\bf  q} = {\bf  k} - {\bf  k} '$. Here $ N _{f} = m k _{F}/2 \pi 
^{2} $ is the density of states at the Fermi surface for equal population 
and $ \bar{\chi } _{0}({\bf  q}) $ is the reduced elementary bubble at 
zero frequency:
\begin{eqnarray}
2 \bar{\chi } _{0}({\bf q}) = 1 + \frac{1}{y} (1 - \frac{y ^{2}}{4} ) 
\ln  \frac{2+y}{2 -y}
\label{eq2}
\end{eqnarray} 
with $ y = q / k _{F}$. The first term in Eq.(1) is the direct term and 
the last one is the indirect interaction due to density fluctuation 
exchange.

The difference in chemical potential $ 2  \bar{ \mu }$ has the effect of 
shifting the kinetic energies of the particle (measured from chemical 
potential) by $ \pm  \bar{ \mu }$. For the pair propagator this is just 
equivalent \cite{larkov} to shift the frequency $ \omega $ by $  \bar{ 
\mu }$. Similarly taking the momenta of the pair members to be $ \pm 
{\bf  k} + {\bf  K}/2$ instead of $ \pm {\bf  k}$ produces a shift by $ 
\pm {\bf k}.{\bf  K}/2m $, leading to an overall shift in frequency by $  
\bar{ \mu } _{k} =  \bar{ \mu } - {\bf k}.{\bf  K}/2m $. At finite 
temperature this means replacing the Matsubara frequency $ \omega 
_{n} = \pi T (2n+1) $ by $ \omega _{n} - i \bar{ \mu } _{k} $. 
On the other hand we expect the modification of the effective interaction 
caused by $ \bar{ \mu }$ to be small, of order $ \bar{ \mu }/ E _{F}$, 
and we can neglect it. This leads us finally to the following gap 
equation at finite temperature : 
\begin{eqnarray}
\Delta _{ {\bf  k}'} = \frac{\lambda T }{ N _{f}} \sum _{n} \int 
\frac{d{\bf  k}}{(2 \pi ) ^{3}} &  & \hspace{45mm}  \nonumber
\end{eqnarray}
\vspace{-5mm}
\begin{eqnarray}
\hspace{13mm}  &  & \frac{\Delta _{ {\bf  k}}}{ \xi _{ {\bf  
k}} ^{2} + \Delta _{ {\bf  k}} ^{2} + (\omega _{n} - i \bar{ 
\mu } _{k}) ^{2}} \ ( 1 +  \frac{1}{2} \ \frac{ \lambda  \bar{\chi } _{0}( 
{\bf  q}) }{1- \lambda \bar{\chi } _{0}({\bf  q})})
\label{eq3}
\end{eqnarray}
where  $\xi _{ {\bf  k}}$ is the kinetic energy measured from the Fermi 
surface for $ \bar{ \mu } = 0 $. Since we are looking for the critical $ 
\bar{ \mu }$ and expect a second ordre phase transition as for the 
FFLO phase, we let $ \Delta _{ {\bf  k}}$ go to zero in the 
denominator. Then, in order to get rid of the cut-off in the frequency 
summation, it is convenient to make use of $ \pi T \sum (1/ | \omega 
_{n} | ) = \lambda  ^{-1} + \ln (T_{c} ^{0}/ T) $, where $ T_{c} 
^{0}$ is the critical temperature for $ \bar{ \mu }= 0 $, $ {\bf  K}=0 $ 
and in the absence of the indirect term in the interaction. When we 
specialize to T $ \rightarrow 0 $, the frequency summation can be 
performed by $ \pi T \sum \rm{sgn} ( \omega _{n}) [ (1/ ( \omega 
_{n} - \dot{\imath} \bar{ \mu } _{k}) - 1/  \omega _{n}] =  \ln (0.88 T 
/  |  \bar{ \mu } _{k} | )  $. In this way we obtain the following equation 
for the critical difference in chemical potential $  \bar{ \mu }$ :
\begin{eqnarray}
\Delta _{ {\bf  k}'} = \int \frac{d \Omega _{k} }{4 \pi } \ \Delta _{ {\bf  
k}} ( 1 + \lambda \ln \frac{ \Delta _{0}}{2  |  \bar{ \mu } _{k} | }) ( 1 
+  \frac{1}{2} \ \frac{ \lambda  \bar{\chi } _{0}( {\bf  q}) }{1- \lambda 
\bar{\chi } _{0}({\bf  q})})
\label{eq4}
\end{eqnarray}
where $  \Delta _{0} $ is the zero temperature gap for $ \bar{ \mu }= 0 
$, $ {\bf  K}=0 $ and in the absence of the indirect term in the 
interaction. When the indirect interaction is not present, $ \Delta _{ {\bf  
k}}$ is independent of $ {\bf  k}$. Then the angular integration is 
easily performed, leading to $ \ln (\Delta _{0}/ 2  \bar{ \mu }) = (1/2) f 
_{0}(x) $, with $ x = K k _{F}/(2m \bar{ \mu })$. Here $ f _{0}(x) = 
\int _{-1} ^{1}du \ \ln | 1+xu | $ which is negative and minimum for $ 
x \approx 1.200 $ ($ x = \coth x $) leading to the standard FFLO result 
$  \bar{ \mu } = 0.754 \Delta _{0} $ .

We consider now the effect of the indirect interaction. Since both $ {\bf  
k}$ and $ {\bf  k}'$ are on the Fermi surface, $q$ goes from 0 to $2k 
_{F}$ and $ \bar{\chi } _{0}( {\bf  q}) $ goes from 1 to 1/2 . 
Although it is not a problem to use numerically the exact expression for 
$ \bar{\chi } _{0}( {\bf  q}) $, one can see that $  \bar{\chi } _{0}( 
{\bf  q}) = 1 - a + a \cos \theta  $, with $ a = 0.25 $ and $ q = 2 k _{F}  
\sin( \theta /2) $ is a quite good approximation. Indeed it gives properly 
the two limits $ q = 0 $ and $ q = 2 k _{F}$. Moreover it corresponds 
to keep only the first two terms in a Legendre polynomials expansion 
with slightly modified coefficients (the exact $ l = 0 $ coefficient is $ ( 1 
+ 2 \ln 2 )/3 = 0.755 $ and the exact coefficient of $ \cos \theta $ is 
0.232 , the higher order terms being fairly small). This approximation 
allows to perform the azimuthal integration analytically, and since our 
calculation is anyway a model calculation it is quite reasonable to make 
this simplification, despite its slight inaccuracy in the vicinity 
of $ \theta = 0 $.

However, before proceeding to the solution of the resulting equation, it 
is interesting to consider first a slightly simplified version which allows 
to carry out the calculation completely explicitely. We merely replace the 
interaction term by the first two terms of its Legendre polynomial 
expansion $ V_{0} + V_{1} \cos \theta $, with $ V_{0} = 0.5 + 0.25 ( 
a \lambda ) ^{-1} \log[1+ 2 a \lambda /(1- \lambda )] $ and $ V_{1} =  
- 1.5 (a \lambda ) ^{-1} + 0.75 ( 1- \lambda + a \lambda ) ({a 
\lambda}) ^{-2} \log[1+ 2 a \lambda /(1- \lambda )] $. In this case the 
solution of Eq.(4) has the form $ \Delta _{{\bf  k}} = 1 + \delta _{1} 
\cos \alpha $ with $  \cos \alpha = \hat{{\bf  k}}. \hat{{\bf  K}}$. Then 
$ L \equiv \ln (2  \bar{ \mu }/ \Delta _{0})$ is solution of the second 
order equation $ (L - 1/ \lambda + 1/ \lambda V_{0}+ 0.5 f _{0})(L  - 
1/ \lambda + 3/ \lambda V_{1}+ 1.5  f _{2}) = 0.75  f _{1} ^{2}$. 
We have set $ f _{n}(x) = \int _{-1} ^{1}du \ u ^{n}\ \ln | 1+xu | $. $ f 
_{1}(x)$ and $ f _{2}(x)$ are both found to be largest for $x$ in the 
range 1.1 - 1.2, and the largest $L$ is also found for the same range of 
$x$ for all values of $ \lambda $. 
\vspace{-4mm}
\begin{figure}
\vbox to 70mm{\hspace{-3mm} \epsfysize=7cm \epsfbox{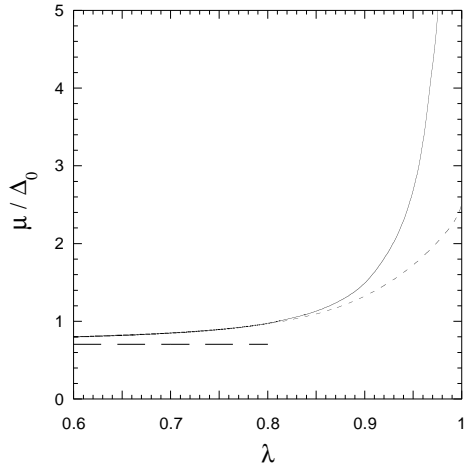} }

\caption{Full line :  location of the second order transition $ \bar{ \mu 
}/  \bar{\Delta} _{0}$ to the normal state from Eq.(4) as a function of $ 
\lambda $. Dashed line : approximate solution with Legendre 
polynomial expansion. Long-dashed line : location of the first order 
transition $  \bar{ \mu }/  \bar{\Delta} _{0}= 1 / \sqrt{2}$.}

\label{figure1}
\end{figure}
Naturally the indirect interaction produces a trivial effect, 
namely the renormalization of the coupling 
constant $ \lambda $ into $ \lambda V_{0}$, producing a 
corresponding change of the critical temperature and of the gap for 
equal population, which goes from  $ \Delta _{0} $ to $ \bar{\Delta} 
_{0} =  \Delta _{0} \exp(1/ \lambda - 1/  \lambda V_{0}) $. This effect 
is simply found by looking at the solution  $ L_{0}$ for $ x = 0 $, that 
is without FFLO phase, since in this case only the isotropic part of the 
interaction is relevant. In the present case this is given by $ L_{0} =  1/ 
\lambda - 1/  \lambda V_{0}$. We are only interested in $ L - L_{0} $ 
which gives the increase in the superfluid domain due to the existence 
of our FFLO phase, in units of $ \bar{\Delta} _{0}$. In Fig.1 we have 
plotted as the dashed line $  \bar{ \mu }/  \bar{\Delta} _{0}$. We see 
that, roughly for $ \lambda < 0.6 $, there is essentially no change with 
respect to the standard FFLO result. And indeed the gap remains essentially 
isotropic in this range. On the other hand, for $ \lambda > 0.6 $, $  
\bar{ \mu }$ increases rapidly and for $  \lambda = 0.9 $ it is almost 
twice the standard FFLO result. It is also quite interesting to consider the 
anisotropy linked to $  \delta _{1} $. As soon as $  \bar{ \mu }$ starts 
to grow with respect to the standard FFLO result, the order parameter gets a 
sizeable anisotropy. For $ \lambda = 0.94 $ we obtain a node on the 
Fermi surface, and for larger $ \lambda  $ a change of sign over the 
Fermi surface with a nodal line.
\vspace{-4mm}
\begin{figure}
\vbox to 65mm{\hspace{-9mm}  \epsfysize=6.5cm \epsfbox{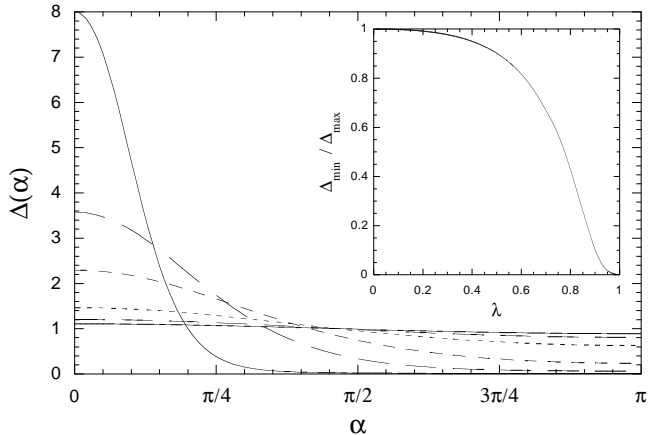}}

\caption{Angular dependence of $ \Delta _{ {\bf  k}}$ for $ \lambda $ 
= 0.6 , 0.7 , 0.8, 0.9 , 0.95  and 0.99 as a function of the angle $ 
\alpha $ between ${\bf  k}$ and $ {\bf  K}$ ($ \Delta $ increases with $ 
\lambda $ for $ \alpha = 0 $). Inset : anisotropy of the order parameter 
as a function of $ \lambda $.}

\label{figure2}
\end{figure}
Let us turn now to the results of the numerical solution of Eq.(4) . They 
are given as the full line on Fig.1 for $  \bar{ \mu }/  \bar{\Delta} 
_{0}$ as a function of $ \lambda $ (as above we have taken into 
account the renormalization of $ \Delta _{0} $ into $  \bar{\Delta} 
_{0}$ by calculating $ L - L_{0} $). For small and intermediate $ 
\lambda $ the result is essentially identical to the result of our simplified 
version. However when $ \lambda $ approaches 1, $  \bar{ \mu }$ 
increases more rapidly. This is easy to understand by looking at Eq.(4). 
In this regime the effective interaction becomes very large for small  $ 
{\bf  q}$ and the log term favors wavevectors nearly along $ {\bf  K}$ 
so it is better to have the gap function $ \Delta _{ {\bf  k}}$ peaked for 
wavevector along $ {\bf  K}$ (see Fig.2). This can be 
seen more explicitely by looking at the simplified form taken by this 
equation when $ \lambda \rightarrow 1 $. Taking into account the 
numerical evidence that in this limit the optimal $  \bar{ \mu }$ is 
obtained for  $ x = 1 $, and making the change of variable $ \hat{{\bf  
k}}. \hat{{\bf  K}} = u (1- \lambda )/a $, we obtain for $ F(u) \equiv  
\Delta _{ {\bf  k}}$:
\begin{eqnarray}
F(v) = - \frac{1}{4a}  \int _{0} ^{ \infty } du \ F(u) \frac{M+ \log u}{ 
\sqrt{1+2(u+v)+(u-v) ^{2}}}
\label{eq5}
\end{eqnarray}
with $ M = L + \log((1- \lambda )/a) - 1 $. This equation can be solved 
numerically. However the solution we are looking for behaves 
approximately as $ \exp(- \mu u)$. Inserting this form into Eq.(5) and 
requiring that $F(v)$ is zero for large $v$ gives $ \log \mu = M - C $ 
(for $a=0.25$), where C is the Euler constant. On the other hand 
requiring $F(0)=1$ and making an asymptotic evaluation of the 
resulting integral gives $ \mu = \exp(-\sqrt{2}) \simeq 0.24 $ which is 
(surprisingly) not much different from the numerical result $ \mu \simeq 
0.3 $. This leads finally to the asymptotic evaluation $  L \approx -1 - 
\log (1- \lambda ) $, which, together with $  L _{0} \approx 1 - 1/  \log 
(0.5/(1- \lambda )) $, is in reasonable agreement with our direct 
solution of Eq.(4) found in Fig.1 . Naturally the divergence of $  \bar{ 
\mu }$ itself is not to be taken seriously since our calculation requires $  
\bar{ \mu } / E _{F} $ to be small anyway.

Coming back to Fig.1 , we see that beyond $ \lambda \approx 0.8 $ the 
results from the full Eq.(4) get larger than those of the simplified 
equation. Hence we find a very large increase of the domain for our
FFLO phase. When we take into account \cite{rc} that the critical 
temperature itself will increase rapidly in this range due to the indirect 
interaction, we see that looking in this region seems quite promising 
experimentally since the overall domain 
for the superfluid phase will be much increased. On the other hand the 
physical properties of this phase will be to a large extent quite different 
from those found for equal population. A first reason is that this phase 
is gapless, just as the standard FFLO phase \cite{ff,larkov}. Next we find an 
important order parameter anisotropy. Indeed we have plotted in Fig.2, 
for various values of $ \lambda $, the angular dependence of $ \Delta _{ 
{\bf  k}}$ with respect to $ {\bf  K}$. As we have mentionned 
already, $ \Delta _{ {\bf  k}}$ gets more concentrated along the  $ 
{\bf  K}$ direction when $ \lambda $ increases. The value of its 
minimum compared to its maximum is also plotted in Fig.2 . We see 
that even for moderate values of $ \lambda $ such as $ \lambda \approx 
0.4 $ the anisotropy is quite sizeable. However when $ \lambda $ is 
further increased the anisotropy becomes ultimately huge to the point that 
$ \Delta _{ {\bf  k}}\approx 0 $ for a very large fraction of the Fermi 
surface. This is a highly unconventional situation  and certainly quite 
unique among BCS superfluids. Another more well-known feature will further 
complicate the matter. We have in our system a degeneracy with respect 
to the direction of $ {\bf  K}$, which will be in general lifted by a 
texture leading to a spatial inhomogeneity, that is yet another symmetry 
breaking, as investigated by LO \cite{larkov} and more recently in Ref. 
\cite{br}. Together with experimental inhomogeneity due to the trap this 
will lead to a remarkably complex physical situation.

In conclusion we have shown that in $^{6}$Li the indirect interaction 
due to density fluctuations exchange will lead to the appearance of a 
new BCS phase with anisotropic order parameter. Near the 
instability threshold, this results in a large increase of the 
superfluid domain as a function of the difference between the atom 
numbers in the two hyperfine states forming the Cooper pairs.

We have much benefited from discussions with A. J. Leggett. We are 
grateful to Y. Castin, J. Dalibard, X. Leyronas, C. Mora and 
C. Salomon for very stimulating conversations.

* Laboratoire associ\'e au Centre National
de la Recherche Scientifique et aux Universit\'es Paris 6 et Paris 7.

\end{multicols}
\end{document}